\documentclass[twocolumn,showpacs,preprintnumbers,amsmath,amssymb]{revtex4}

\usepackage{graphicx}
\usepackage{dcolumn}
\usepackage{bm}
\usepackage{amsmath, amsthm, amssymb}

\newtheorem{theorem}{Theorem}[section]

\begin{document}


\title{Completely Positive Markovian Quantum Dynamics in the Weak-Coupling Limit}

\author{David Taj}
 \email{david.taj@gmail.com}
\author{Fausto Rossi}%
 \affiliation{ Dept. Physics, Politecnico di Torino, C.so Duca degli Abruzzi 24, 10129, Torino, Italy
 }%



\begin{abstract}
We obtain new types of exponential decay laws for solutions of
density-matrix master equations in the weak-coupling limit: after
comparing with results already present in the literature and
developing the necessary techniques, we study the crucial aspect
of complete positivity under fairly general conditions. We propose
a new type of time average that guarantees complete positivity and
approximates, in markovian fashion, the exact dynamics for a
plethora of physical applications, no matter which are the
spectral properties of the subsystem, or its dimensions. We shall
comment on some interesting examples, like a new Quantum version
of the celebrated Fermi's Golden Rule and some recently proposed
entangling projections.
\end{abstract}

\pacs{
05.30.-d
03.65.Yz,
72.10.Bg,
85.35.-p,
02.30.Sa,
02.30.Tb, }

\maketitle

\section*{Introduction}
After the pioneering works in 1974 and 1976 by
Davies~(\cite{davies1,davies2}), a huge amount of physical
information about the irreversibility and the evolution of open
quantum mechanical systems has been gained. Alicky~\cite{alicki}
showed in 1977 that these efforts were deeply connected to the
celebrated "Fermi's Golden Rule", that now had become
mathematically consistent. The conceptual importance of these
works is clearly not only academic, as the need for a better
understanding of irreversible processes has never been more
urgent. Today, so many nanotechnologies are pushing devices
towards limits where neither quantum phase coherence, nor
dissipation/dephasing, can be neglected~\cite{SL,QD,RMP}. Many
attempts to improve the theory have been made since then (see for
example~\cite{breuer,open}), but despite the compelling need, no
substantial, fundamental progress, directly applicable to nowadays
technologies, has been made so far.

To be more specific, the problem is to understand the dynamics of
a subsystem of interest, when the global system is fully coherent.
In many cases, the dynamics of the global system can be splitted
into a part that leaves the subsystem invariant, plus an
interaction between the subsystem and the remaining
"uninteresting" degrees of freedom. The problem then arises
whether or not the subsystem can be given a markovian, possibly
dissipative, dynamics as a consequence of its interaction with
those degrees of freedom. This is generally impossible of course,
but it has been shown since the '70s that a positive answer can be
given when the amount of uninteresting degrees of freedom is huge,
and the interaction is made small. This last condition is referred
to as the "weak-coupling limit".

In~\cite{davies1} the author was able to give a solid physical
model of a discrete "atom" (system A) interacting with a fermionic
particle reservoir at thermal equilibrium (environment B). In that
case, the subsystem was made of unentangled pairs of atomic states
(also referred to as "density matrices") and a fermionic thermal
equilibrium state, while all the entangled pairs just constituted
the remaining, uninteresting, degrees of freedom. The model was of
high conceptual importance, as the markovian (and dissipative)
dynamics for the subsystem was shown to guarantee the state
positivity at all times. This fact gave just enough internal
robustness to the model as to be of invaluable practical use, for
at any time, the subsystem evolution could be given a strong
physical meaning. But the system A had to be finite dimensional,
or at least, its unperturbed hamiltonian was forced to have
discrete spectrum.

Since then, much research has been (and currently is) focusing on
the so called "projection techniques". The name "projection" comes
indeed from the fact that, as explained brilliantly in
\cite{breuerpra}, a certain projection superoperator $P_0$ is
introduced: $P_0$ "is the mathematical expression for the idea of
the elimination of degrees of freedom from the complete
description of the states of the total system". The effort is here
to understand what could be the basic requirement for choosing the
subsystem/$P_0$ in order to guarantee a positive subsystem
dynamics. Much information on the general properties of some
bipartite subsystems has thus been gained
\cite{breuerpra,vacchini,budini}, which is relevant in the field
of (possibly entangled) Open Quantum Systems, where one studies
the dynamics of a "small" system A interacting with a "big"
environment B (not necessarily at thermal equilibrium). Usually,
one supposes to be given a generic "phenomenological" markovian
and dissipative dynamics for the global system, and writes down a
(markovian dissipative) master equation for the subsystem. But, as
we shall see in the following, the strong interplay between the
projection (i.e. subsystem) and the unperturbed global dynamics,
severely restricts the possibilities for $P_0$, or on the other
side, for the global system dynamics. For this reason, although
these models are much relevant to nowadays technologies, they all
still suffer from the original restriction, that system A should
be finite dimensional, or at least it should have discrete
spectral properties, just like the original Davies' model
\cite{davies1}. We have to say here that alternative important
methods have been addressed since a long time (see for example
\cite{davies3}), and still are (see~\cite{budini}), which are
based on stochastic techniques: these models are of course of
enormous practical use, but since some randomness has to be put by
hand, they are somehow of less fundamental nature.

Unfortunately, the possibility to obtain a "completely positive
markovian subsystem dynamics" ---also referred to as Quantum
Dynamical Semigroups, or QDSs \cite{lindblad}--- through the
weak-coupling limit procedure, becomes no more available when the
system is infinite dimensional. In that case indeed Davies showed
that a markovian approximation, for the subsystem exact projected
dynamics, could still be achieved \cite{davies2}, but positivity
could not be shown in general, if not up to finite times: not even
for the old and "safe" partial tracing over the thermal bath's
degrees of freedom. Although more general, the theory did not
share the enormous success of the previous one among physicists,
precisely because of this serious limitation. For example, all of
the steady state analysis became, physically, completely
meaningless.

In this work we follow the path set by Davies, in supposing the
global system is undergoing a fully coherent (hamiltonian)
dynamics, and propose a new master equation for the subsystem, on
the basis of time-symmetry considerations. At this point, we will
introduce a new object, called "completed collision time", which
represents a safe overestimation of the subsystem relaxation
times, as well as a safe underestimation of the global system
recurrence time. This will be a key physical step, that is amply
justified from the following well known fact (see for example
\cite{lindblad,FGR,cohen} among many others): the "semigroup" or
markovian approximation is physically meaningful only when neither
we observe the subsystem at short times (where probabilities of
excited states have a parabolic time dependence), nor we observe
it long enough to see Poincar\'e recurrence cycles. Indeed these
cycles are present, and explicitly appear in the subsystem exact
dynamics, whenever the global dynamics is hamiltonian. This time
will scale with the coupling constant, and will serve us to
"dynamically diagonalize" our time-symmetrized master equation,
thereby eliminating in symmetric fashion most of the off-diagonal
terms in the generator. We will then be able to show that our
master equation will correctly describe the exact subsystem
dynamics in the weak-coupling limit, it will be defined
irrespective of the subsystem dimensions or spectral properties,
and it will guarantee positivity for the dynamics, for a plethora
of different projections/subsystems. The number of possible
projections is indeed huge enough to incorporate the partial trace
over the bath, together with some more recently proposed
"entangling" projections \cite{breuerpra,vacchini}, and some
others by us, that furnish a new Quantum version for the Fermi's
Golden Rule (QFGR).

This deserves a comment: there are many proposed quantum
generalization of the well known Fermi's Golden Rule
\cite{open,FGR}, which are robust and physically and
mathematically meaningful. All these generalizations consider a
bipartite system (this in fact is the only way to propose the
standard Fermi's Golden Rule in mathematical terms,
see~\cite{alicki}): nevertheless, the original idea by Fermi,
although mathematically ill-posed, did not come from any sort of
(rather sophisticated) projection on some bipartite system. Stated
in modern terms, Fermi's idea was rather to take a global system
and project on the space of density matrices, that are diagonal in
the basis of the unperturbed hamiltonian. His motivations referred
to the fact that the system eigenvalues are "robust" against
dissipation, and thus constitute the relevant degrees of freedom.
Just along these lines, no environment will be present in our
model of the QFGR, and the relevant degrees of freedom will be
"box-diagonal" density matrices for a "box-diagonal" unperturbed
hamiltonian. The possibility to obtain a semigroup (a dissipative
and markovian dynamics) comes exactly by the fact that the
unperturbed hamiltonian has continuous spectrum, thus conferring
our QFGR version an autonomous relevance.

It is worth noting that the explicit form for our master equation
in this case will share some very similar and interesting formal
properties with some recently appeared master equations coming
from the theory of Entangled Open Quantum
Systems~\cite{breuer,vacchini}. These, as ours, are linear
homogeneous equations for a "composed" density matrix, which turn
into a linear homogeneous system of equations for each component.
Needless to say, these kind of equations are currently showing an
enormous potential, and will be shown here to be taken into
account by our general theory. But contrary to our QFGR, these
models are all based on bipartite systems.

Two final introductory remarks. First, we have to say that,
contrary to what is sometimes believed, a "diagonal" superoperator
\emph{does not} imply that the dynamics is free from quantum
effects: first of all, "diagonal" refers to some basis, say for
example that induced by some unperturbed hamiltonian $H_0$. Then,
talking in Heisenberg picture, it is true that a diagonal
generator will leave a subset of observables ---those commuting
with $H_0$--- invariant (and this would be a remarkable
property!). But the subsystem \emph{needs not} be formed only by
such a subset, and nontrivial commutation relations could well be
observed within the subsystem evolution, even for "diagonal"
generators. With respect to this, our theory is highly
non-trivial, as it fully keeps track of quantum effects and does
not, to any extent, constitute a sort of semiclassical limit. This
is at variance, for example, with many widely used Green functions
truncation schemes, where one takes "$n$-point" quantum
correlations functions into account, but only up to some finite
$n$ (from this point of view, the weak-coupling and the Green
function theories are applied in completely different contests, a
fact that is often misunderstood). Rather, the fact that a
generator is diagonal in the unperturbed basis will be shown to be
strictly linked with its algebraic properties, like that of the
positivity of its generated dynamics.

Finally, we have to say that we have made explicit use of Hilbert
spaces instead of the (much more natural environment of) Banach
spaces. This is justified by the fact that the Banach space
$\mathcal{T}(\mathcal{H})$ of trace class operators on a Hilbert
space $\mathcal H$ can be equipped with a scalar product, and the
resulting completion is a Hilbert space (that of of
Hilbert-Schmidt operators on $\mathcal{H}$). Our choice only comes
from simplicity in the proofs involved, as we can make use a huge
amount of spectral theory, but of course is still a rather serious
limitation. Nevertheless we feel that our message could be
extended to Banach spaces, and we are currently obtaining
promising results in this direction. This indicates that the
physical content of what we discuss here should survive the
mathematical formalism, substantially unmodified. We also want to
assure the less mathematically inclined (or more physically
motivated) reader, that our theorems, although mathematical
statements, will be critically discussed within a fully
self-contained physical contest, that will offer, so to say, the
physical arguments that led us to the proofs. These will instead
be reported in the appendices, and are directed to the more
mathematically oriented readers.

The paper is organized as follows: in the next Section we begin
with a warm up to clarify notations and state some physical
example to motivate our theory, which we then present. In Section
\ref{sec:symm} we discuss a new time symmetrized markovian
generator and make contact with some preliminary results by one of
us, reported in~\cite{fausto}; in Section \ref{sec:time} we
propose our dynamically time averaged generator; finally in
section \ref{sec:positivity} we find some fairly general class of
projections that make our generator into a Lindblad
form~\cite{lindblad}, thus obtaining a generator of a (completely
positive markovian) Quantum Dynamical Semigroup. We will also
discuss some (already cited) important applications, such as our
Quantum version of the Fermi's Golden Rule and some recently
appeared entangling projection~\cite{breuerpra}. Then we shall
conclude, and report our proofs in the Appendices.

\section{General Framework}
Let us briefly run through some well known examples, that will
serve us to better state what we aim at, clarify notation and
motivate the general theory.
\subsection{Partial Trace}\label{sec:partial_trace}
This is also referred to as "tracing away the bath degrees of
freedom": let $\mathcal{H}=\mathcal{H}_A\otimes\mathcal{H}_B$ be
the tensor product of two Hilbert spaces, representing a coupled
system-environment pair. Let ${\mathcal{B}}$ be the space of
trace-class operators on $\mathcal{H}$ and ${\mathcal{B}}_0$ the
space of trace-class operators on $\mathcal{H}_A$. These are
Banach spaces, but they can be completed with the scalar product
$\langle A,B\rangle=\mbox{Tr}(A^\dagger B)$ to the Hilbert spaces
(respectively $\mathcal{H.S.}(\mathcal{H})$ and
$\mathcal{H.S.}(\mathcal{H}_A)$) of Hilbert-Schmidt operators on
the respective Hilbert spaces: we shall always work within these
spaces, which we shall denote with the same letters $\mathcal{B}$
and $\mathcal{B}_0$. Now define the linear map, called the
"partial trace",
$\overline{P}_0:{\mathcal{B}}\rightarrow{\mathcal{B}}_0$ uniquely
determined by
\begin{equation}
\textit{Tr}(\overline{P}_0(\rho)X_A)=\textit{Tr}(\rho(X_A\otimes
1_B))
\end{equation}
for arbitrary $\rho\in {\mathcal{B}}$ and bounded operator $X_A$
on $\mathcal{H}_A$. If a state
$\sigma\in\mathcal{H.S.}(\mathcal{H}_B)$ is given on
$\mathcal{H}_B$, then $P_0\rho=\overline{P}_0(\rho)\otimes \sigma$
is a projection in ${\mathcal{B}}$ with image ${\mathcal{B}}_0$,
and is called "the partial trace". Then we have a direct sum
decomposition
\begin{equation}
\mathcal{B}=\mathcal{B}_0\oplus\mathcal{B}_1 \;,
\end{equation}
where $\mathcal{B}_i=P_i(\mathcal{B})$ and we have put
$P_1=1-P_0$.

Now suppose we are given a hamiltonian of the form $H=H_0+\lambda
H_I$ with $H_0=H_A\otimes 1 +1\otimes H_B$ and $H_I=\sum_n
\Phi_n\otimes \Psi_n$ (take all operators hermitian). Call $Z$ the
generator of the one parameter group $U_t$ of unitarities, on the
global system $\mathcal{B}$, defined by the von Newmann equation
\begin{equation}
\partial_t\rho=-i[H_0,\rho]=Z\rho \;.
\end{equation}
Then we have
\begin{equation}
U_t \rho = e^{Z t}\rho = e^{-i H_0 t}\, \rho\, e^{i H_0 t} \;,
\end{equation}
And one obtains the all important (superoperator) commutation
property
\begin{equation}
[Z,P_0]=0\;.
\end{equation}
This is telling us that the relevant degrees of freedom (the
subsystem $\mathcal{B}_0=P_0(\mathcal{B})$ \emph{do not} interact
with the irrelevant ones (the "superpolarization space"
$\mathcal{B}_1$) by means of the unperturbed dynamics $U_t$. This
is a fundamental assumption, that intrinsically links the
projection $P_0$ to the way the dynamics is splitted into a
perturbation and an unperturbed one, and that should always be
checked to avoid drastic mistakes: if this hypothesis is not
verified, the whole theory of the weak-coupling limit simply
breaks down.

At this point we introduce the perturbation superoperator
$A\rho=-i[H_I,\rho]$ so that we can state with precision what we
aim at: we simply would like to know something about the
projection at current time $t$ of the unitary group describing the
exact global hamiltonian dynamics. That is, we want to study
\begin{equation}
\rho_t= P_0\, e^{(Z+\lambda A)t}P_0\rho=W^\lambda_t\rho
\end{equation}
for a given initial condition $\rho$ in $\mathcal{B}$. In passing,
we have defined the subsystem evolution operator
$W^\lambda_t=P_0\, e^{(Z+\lambda A)t}P_0$ (which of course is
\emph{not} a one parameter group of unitary transformations).
\subsection{Diagonal Matrices in a Closed Quantum System}\label{sec:fermi}
This is close to the philosophy followed by Fermi. Let
$\mathcal{H}=\mathbb{C}^n$ be the $n$-dimensional Hilbert space
over the complex field: then trace class operators are just
$n\times n$ matrices. Let these matrices form the total state
space $\mathcal{B}$, and let a hamiltonian $H_0$ be given. Let
$|\alpha\rangle$ denote its eigenbasis, denote with
$\mathcal{B}_0$ the space of diagonal elements of $\mathcal{B}$ in
such basis, and define the projection $P_0$ on $B_0$ by
\begin{equation}
P_0(\rho)=\sum_{\alpha=1}^n |\alpha\rangle\langle \alpha |\rho
|\alpha\rangle\langle\alpha| \;.
\end{equation}
Then as before one defines $Z$ through $H_0$ and gives a
perturbation $A$ through some hermitian $H_I$. Obviously
$[Z,P_0]=0$, and the diagonal part of the evolution at time $t$ of
the initial state $\rho$ in $\mathcal{B}$ is $\rho_t=P_0\,
e^{(Z+\lambda A)t}P_0\rho=W^\lambda_t\rho$.

\subsection{General Theory}\label{sec:general}
We are ready to state the general theory: we suppose that $P_0$ is
a (not necessarily orthogonal) projection on a Hilbert space
$\mathcal{B}$, put $P_1=1-P_0$ and $\mathcal{B}_i=P_i\mathcal{B}$,
so that
\begin{equation}
\mathcal{B}=\mathcal{B}_0\oplus\mathcal{B}_1.
\end{equation}
We suppose that $Z$ is the (skew-adjoint) generator of a strongly
continuous one parameter group of unitarities $U_t$ on
$\mathcal{B}$ with
\begin{equation}
U_t P_0=P_0 U_t
\end{equation}
for all $t\in\mathbb{R}$, or equivalently
\begin{equation}
[Z,P_0]=0
\end{equation}
and put $Z_i=P_i Z$. We suppose that $A$ is a bounded perturbation
of $Z$ and put $A_{ij} = P_iAP_j$. We shall always suppose, for
simplicity, that no first order term are present for the
subsystem. That is, we shall take $A_{00}=0$ all throughout. We
let $U^\lambda_t$ be the one parameter group generated by $Z +
\lambda A_{11}$ and let $V_t^\lambda$ be the one parameter group
generated by $(Z+\lambda A)$. Then putting
\begin{equation}
W^\lambda_t=P_0 V^\lambda_t P_0
\end{equation}
one obtains an exact and closed equation for the projected
dynamics:
\begin{equation}\label{eq:exact}
W^\lambda_t=U_t+\lambda^2 \int_{s=0}^t \int_{u=0}^s
U_{t-s}A_{01}U^\lambda_{s-u} A_{10} W^\lambda_u \; ds du.
\end{equation}
This is nothing but the integrated form of the well known master
equation constructed by Nakajima, Prigogine, Resibois, and Zwanzig
\cite{naka,zwanzig}.

It is a good moment here to note that, strictly speaking, $P_0$
\emph{does not coincide} with the physical subsystem: rather, the
subsystem should be identified with the \emph{image} of $P_0$,
that is with $\mathcal{B}_0$. We'll see that dissipative effects
will be depending also upon the complete features of $P_0$ itself,
so \emph{not only the relevant degrees of freedom $\mathcal{B}_0$
determine the semigroup dynamics: also the way we look at them,
with $P_0$, does influence the evolution.} With this in mind, we
shall use the terms "subsystem" or "$P_0$" equivalently, when no
there is no room for misinterpretations.

\section{Time-symmetric markovian approximation}\label{sec:symm}
Changing variables to $x=s-u$, $\sigma=\lambda^2 u$ and
introducing the time rescaled interaction picture dynamics
$W^{\lambda,i}_\tau=U_{-\lambda^{-2}\tau}
W^\lambda_{\lambda^{-2}\tau}$, one is led to \cite{davies2}:
\begin{equation}\label{eq:davies_int}
W^{\lambda,i}_\tau=1+ \int_{\sigma=0}^\tau U_{-\lambda^{-2}\sigma}
K(\lambda,\tau-\sigma)U_{\lambda^{-2}\sigma} W^{\lambda,i}_\sigma
d\sigma,
\end{equation}
where
\begin{equation}
K(\lambda,\tau)=\int_0^{\lambda^{-2}\tau} U_{-x}A_{01}U^\lambda_x
A_{10}\; dx.
\end{equation}
This form separates an "interacting" and "slowly varying" part
$K(\lambda,\tau)$ from the "rapidly oscillating" free-evolution
$U_{\lambda^{-2}\sigma}$ to second order in the coupling constant
$\lambda$. Now in the weak-coupling limit $\lambda\rightarrow 0$,
the slowly varying integral kernel $K(\lambda,\tau)$ converges to
\begin{equation}
K_D=\int_0^{\infty} U_{-x}A_{01}U_x A_{10}\; dx \;,
\end{equation}
where $K_D$ is the celebrated Davies' superoperator. Substituting
$K_D\sim K(\lambda,\tau)$ in (\ref{eq:davies_int}) and moving back
to the "Schr\"odinger picture" we obtain the markovian
approximation for our subsystem dynamics
\begin{equation}
{W}^\lambda_{t}\sim \overline{W}^\lambda_{t}=e^{(Z_0+\lambda^2
K_D)t} \;.
\end{equation}
Indeed in \cite{davies2} an important theorem shows that under
reasonable and general conditions the approximation holds in the
weak-coupling limit, up to $\lambda^{-2}$-rescaled times,
independently of the subsystem dimensions or spectral properties.
Unfortunately, although $K_D$ is defined irrespective of the
subsystem spectral properties, it does not guarantee positivity of
the generated dynamics.

However, uniqueness of $K_D$, that is, of the semigroup
approximation, is by no means to be expected: already in case
$\mathcal{B}_0$ is finite dimensional (or more generally when
$Z_0$ has discrete spectrum) one can define a time average of
$K_D$ (which takes the diagonal part of $K_D$ in the $Z$
"super"-eigenbasis), and show that the generated dynamics is still
asymptotic to the exact one in the weak-limit. In the infinite
dimensional case this averaging map is not generally defined, but
still no "physical" argument would imply uniqueness.

With respect to this, we shall now propose a new time-symmetrized
markov approximation: changing variable in eq. (\ref{eq:exact}) to
\begin{equation}
\left\{
\begin{array}{rcl}
\sigma &=& {\lambda^2\over 2}(s+u)  \\
r&=&s-u,
\end{array}
\right.
\end{equation}
and working again in the time-rescaled interacion picture, we
obtain
\begin{widetext}
\begin{equation}\label{eq:rossi_int}
W^{\lambda,i}_\tau=1+ \int_{\sigma=0}^\tau U_{-\lambda^{-2}\sigma}
\int_{r=0}^{\lambda^{-2}g(\tau,\sigma)} U_{-{r\over
2}}A_{01}U^\lambda_r A_{01} U_{-{r\over 2}} U_{\lambda^{-2}\sigma}
W^{\lambda,i}_{(\sigma+\lambda^2 {r\over 2})} d\sigma dr.
\end{equation}
\end{widetext}
with $g(\tau,\sigma)=|\tau/2-|\sigma-\tau/2||$. Now we argue as
follows: by direct inspection we see that in the weak-coupling
limit $\lambda\rightarrow 0$, the function
$\lambda^{-2}g(\tau,\sigma)$, defining the integration domain for
the $r$-variable, converges to
\begin{equation}
\lambda^{-2}g(\tau,\sigma) \rightarrow +\infty
\end{equation}
for each $\sigma\in(0,\tau)$. But at the same time, the
time-rescaled interaction picture dynamics moves slowly with
respect to the $r$-variable (again when $\lambda\sim 0$), and this
fact makes us guess that we can further approximate
\begin{equation}\label{eq:rossi_approx}
W^{\lambda,i}_{(\sigma+\lambda^2 {r\over 2})}\sim
W^{\lambda,i}_{\sigma}
\end{equation}
in the dynamics (\ref{eq:rossi_int}). Then, the "freezed" integral
in the $r$-variable factors, and our integral equation can then be
easily seen to turn into the markovian subsystem evolution
\begin{equation}
\widetilde{W}^\lambda_{t}=e^{(Z_0+\lambda^2 K_R)t},
\end{equation}
with
\begin{equation}\label{eq:K_rossi}
K_R=\int_0^{\infty} U_{-{x\over 2}}A_{01}U_x A_{10}U_{-{x\over
2}}\; dx.
\end{equation}
Here $K_R$ is the superoperatorial and projected version of what
Rossi proposes in \cite{fausto}, that also includes the second
order energy renormalization effects.

In fact, let us consider $Z\rho=-i[H_0,\rho]$ and
$A\rho=-i[H',\rho]$ for self-adjoint $H_0$ and (bounded) $H'$ on a
Hilbert space $\mathcal{H}$. Denoting $A(t)=U_{-t}AU_t$ and
$H'(t)=e^{-i H_0 t} H' e^{i H_0 t}$, we factor $K_R=P_0
\widetilde{K}_R P_0 $ where,
\begin{eqnarray}\label{eq:explicit_rossi}
\widetilde{K}_R \rho &=& \int_0^\infty dx\; A\left({x\over 2}\right)A\left(-{x\over 2}\right)\rho\nonumber\\
&=&-\int_0^\infty dx\; \left[H'\left(-{x\over
2}\right),\left[H'\left({x\over 2}\right),\rho\right]\right],
\end{eqnarray}
which in fact makes contact with the explicit form given in
\cite{fausto}.

Clearly, starting from the interactions $A_{ij}$, the
superoperator $K_R$ is built in a much more time-symmetric fashion
than $K_D$, as can be directly seen by inspection of
(\ref{eq:K_rossi}). Of course, all our arguments here are but a
physical motivation, and we must prove that our markovian
approximation indeed is consistent with the exact dynamics, at
least up to $\lambda^{-2}$-rescaled times. From what we have
stated here, we see that the main difficulty is to justify tha
fact that the convergence $W^{\lambda,i}_{(\sigma+\lambda^2
{r\over 2})}\rightarrow W^{\lambda,i}_{\sigma}$ \emph{is faster}
than the convergence $g(\tau,\sigma) \rightarrow \infty$, when
$\lambda\rightarrow 0$, or alternatively, that the subsystem
evolves \emph{slowly} compared to a $\lambda^{-2}$-time scale, in
the weak-coupling limit.

This problem presents surprising difficulties when $\mathcal{B}$
is a general Banach space, but as we have said, it can be attacked
with a spectral analysis when we suppose $\mathcal{B}$ to be a
Hilbert space, and the final answer is positive.

The idea is precisely to compare our markov semigroup
approximation with that of Davies (outlined before), and show
compatibility when $\lambda\rightarrow 0$. Compatibility of our
semigroup with the exact projected dynamics would then follow
automatically, thanks to the work by Davies in \cite{davies2}.
What we do is to estimate the greatest difference between the two
semigroups at times from zero up to time
$\lambda^{-2}\overline{\tau}$, where $\overline{\tau}$ is some
initially, arbitrarily chosen, positive reference time. Then we
show that this difference goes to zero in the weak-coupling limit.
So we state our
\begin{theorem}\label{th:symmetric}
Suppose that
\begin{equation}\label{eq:no_oscillation}
\int_0^\infty \|A_{01}U_t A_{10}\| \, dt < \infty
\end{equation}

Then for every $\overline{\tau}>0$
\begin{equation}
\lim_{\lambda\rightarrow 0} \left\{ \sup_{0\leq t\leq
\lambda^{-2}\overline{\tau}}
\|\overline{W}^\lambda_t-\widetilde{W}^\lambda_t\|\right\}=0.
\end{equation}
\end{theorem}

We defer the proof to Appendix A, for the mathematically inclined
readers, while we now think we have completely justified and
commented the theorem from a physical point of view. But we still
have to understand the physical meaning of the all-important
hypothesis (\ref{eq:no_oscillation}): the integrand $\|A_{01}U_t
A_{10}\|$ has a very simple interpretation as the
"subsystem-to-superpolarization space super-correlation function".
In fact, reading it from right to left, we start in our subsystem
$\mathcal{B}_0$, interact and go into the "super-polarization
space" $\mathcal{B}_1$, evolve with the unperturbed $U_t$, and
finally interact a second time to land back into our subsystem
$\mathcal{B}_0$. The norm is just the mathematical tool to
evaluate how big such a "super"-amplitude is. Our hypothesis is
thus telling us that everything works fine \emph{if these
correlations fall faster than} $1/t$ as the time $t$ increases.
That is, the super-polarization space $\mathcal{B}_1$ is so huge,
that even though the global dynamics is fully coherent and
hamiltonian, \emph{information leaves the subsystem very fast, at
least from a given time on}. This hypothesis obviously excludes
from a markovian approximation any finite (and even countably
infinite) dimensional global system: in this case, the only things
that one is going to see in its subsystems are nothing but some
projected forms of \emph{never-damping} Block oscillations. This
explains the need to use an infinite-dimensional bath with
continuous spectrum in Subsection \ref{sec:partial_trace}, and the
impossibility to observe dissipation in Subsection
\ref{sec:fermi}.

Unfortunately we are not finished yet, as the superoperator $K_R$
\emph{does not} generate a positive semigroup. But the effort we
have done so far in symmetrizing the well known usual markovian
approximation $K_D$ will be completely paid in a moment: the idea
is that $K_R$ is, as $K_D$, highly non-diagonal in the $Z_0$
eigenbasis. But contrary to $K_D$, the upper triangular part of
$K_R$ in such a basis perfectly mirrors its lower triangular part.
So if now we could somehow remove the off diagonal terms in a
simple symmetric fashion, we should be done in obtaining a fully
diagonal superoperator. Needless to say in fact, a diagonal
superoperator must be intrinsically linked with positivity
features of its generated dynamics. But the term "diagonal part"
must be handled with caution, as strictly speaking, it is a
measure zero sector when the subsystem hamiltonian spectrum is
continuous. We shall thus here propose a "dynamic" version of the
averaging map in \cite{davies1}, that extracts a diagonal "cloud"
from $K_R$ whose width goes to zero only in the limit
$\lambda\rightarrow 0$. Let us see how.

\section{Spectral Diagonalization and Completed Collision Time}\label{sec:time}
In \cite{davies1} an averaging map is introduced: for an operator
$K: \mathcal{B}_0 \rightarrow \mathcal{B}_0$, we put
\begin{equation}\label{eq:spavdav}
K^\natural=\lim_{T\rightarrow \infty} {1\over 2T}\int_{-T}^T dq\;
U_{-q} K U_q
\end{equation}
whenever the right hand side is defined. Then Davies shows in
\cite{davies2} that if $\mathcal{B}_0$ is finite dimensional, then
the operation $\natural$ is well defined and completely recovers
the exact subsystem dynamics in the weak limit.

Unfortunately, as said before, the spectral average $\natural$ is
not generally defined when the system hamiltonian has continuous
spectrum. We thus introduce a new type of spectral averaging: we
note that the integral over time in eq.(\ref{eq:spavdav}) can be
extended to the entire real line by putting a $T$-related damping
term inside the integral:
\begin{equation}\label{eq:spavt1}
K^\natural=\lim_{T\rightarrow +\infty} {1\over \sqrt\pi T}
\int_{-\infty}^\infty dq\; e^{-q^2/T^2} U_{-q} K U_q.
\end{equation}
The factor $\sqrt\pi$ is needed for the two definitions of
$\natural$ to coincide when $\mathcal{B}_0$ is finite dimensional:
then the spectral decomposition
\begin{equation}
Z_0=\sum_\alpha i \omega_\alpha Q_\alpha
\end{equation}
gives
\begin{eqnarray}
K^\natural &=& \lim_{T\rightarrow +\infty} {1\over 2 T}\int_{-T}^T dq\; e^{i(\omega_\alpha-\omega_\beta)q}\sum_{\alpha\beta} Q_\alpha K Q_\beta \nonumber \\
&=&\lim_{T\rightarrow +\infty} {1\over \sqrt\pi T} \int_{-\infty}^\infty dq\; e^{-q^2/T^2}e^{i(\omega_\alpha-\omega_\beta)q}\sum_{\alpha\beta} Q_\alpha K Q_\beta \nonumber\\
&=&\sum_\alpha Q_\alpha K Q_\alpha
\end{eqnarray}
for both definitions (\ref{eq:spavdav}) and (\ref{eq:spavt1}).

We may ask ourselves about the physical origin of this new object
$T$: it appears that the higher its value, the finer is our
ability to separate the system characteristic frequencies. This is
exactly what happens when we tune our radio on a given frequency:
the more we wait, the more precise we are in counting the
transmitted oscillations, and thus the finer we are in resolving
the wanted frequency. So this $T$ looks like a subsystem
observation time: we have to wait an infinite amount of time to
resolve the system frequencies \emph{exactly}, but it is everyday
experience that we only have a \emph{finite} (but long enough)
observation time at our disposal. The idea is then to make use of
such observation time $T$ to compare with the subsystem relaxation
times, and the global system recurrence times. It could be
referred to as an "observation time" (but also, as we shall see
later on, "completed collision time"): it is somewhat proportional
to our instrumental uncertainty, or lack of precision, on the
energy measurements we perform, since we study the system
evolution only up to finite times. This way of understanding the
origin of $T$ is consistent with the fact that when $Z_0$ has
discrete spectrum, then the averaging map (\ref{eq:spavdav}) is
well defined. In fact, while we raise $T$, thus raising our energy
measurements precision, we meet a finite value for $T$ over which
we are precise enough to unambiguously distinguish among the
different system frequencies. This is not the case when the system
frequencies (those of $Z_0$) form a continuous set, as there's no
finite value for $T$ beyond which the system energy levels are
distinguishable. In any case, our precision in measuring them gets
higher as we're able to increase $T$.

After having so understood the physical origin of $T$ we could as
well introduce a $T$-smoothed version of $\widetilde{K}_R$ (see
eq.(\ref{eq:explicit_rossi}) for notation),
\begin{equation}
\widetilde{K}^T=\int_0^\infty dx\; e^{-\left({x\over
2}\right)^2/T^2}\: A\left({x\over 2}\right)A\left(-{x\over
2}\right),
\end{equation}
which of course converges to the previous one in the limit
$T\rightarrow \infty$, and put $K^T=P_0 \widetilde{K}^T P_0$. Then
we have yet another expression for the spectral average of $K_R$:
\begin{equation}
K_R^\natural = \lim_{T\rightarrow +\infty} K_T.
\end{equation}
where we have defined
\begin{eqnarray}\label{eq:kappat}
&&K_T = {1\over \sqrt\pi T} \int_{-\infty}^\infty dq\; e^{- q^2/T^2} U_{-q} K^T U_q \nonumber \\
&&={1\over \sqrt\pi T} \int_{-\infty}^\infty\!\! dt_1 e^{-{t_1^2\over 2 T^2}}A_{01}(t_1)\!\!\int_{-\infty}^{t_1}\!\! dt_2\;  e^{-{t_2^2\over 2 T^2}}  A_{10}(t_2) \nonumber \\
\end{eqnarray}
and in the last line we have changed variable to $t_1=q+x/2$ and
$t_2=q-x/2$. This last expression deserves some comments: first,
it is probably the most symmetric thing that one could imagine to
build from basic objects ---such as $P_0$, $Z$ and $A$--- starting
from the Nakajima-Zwanzig memory term in (\ref{eq:exact}). Second,
unlike our preliminary $K_R$ in (\ref{eq:K_rossi}), $K_T$ gets
more and more diagonal in the $Z$ eigenbasis, as is evident by the
procedure we use to obtain it. Third, the gaussian smoothing
---sometimes referred to as homogeneous broadening--- seems not to
be there by chance, or in other words, an exponential decay
---also referred to as inhomogeneous broadening--- would
\emph{not} allow one to so easily pass from the first line to the
second in (\ref{eq:kappat}). In fact, the gaussian is the only
distribution $\Phi$, such that
\begin{equation}
\Phi(q^2+(x/2)^2)=\Phi((t_1^2+t_2^2)/2)
\end{equation}
for $t_1=q+x/2$ and $t_2=q-x/2$.

Now the important thing to note is that the limit $T\rightarrow 0$
is well defined, as said above, only when $\mathcal{B}_0$ is
finite dimensional, or more generally when $Z_0$ has discrete
spectrum, but, for finite values of $T$, $K_T$ and
$\widetilde{K}_T$ are always well defined, no matter which are the
spectral properties of $Z_0$ nor the dimension of $\mathcal{B}_0$,
and moreover we can write $\widetilde{K}_T$ explicitly as:
\begin{eqnarray}
\widetilde{K}_T\rho &=& {1\over \sqrt\pi T} \!\!\int_{-\infty}^\infty\!\! dt_1 \!\!\int_{-\infty}^{t_1}\!\! dt_2\; e^{-{t_1^2+t_2^2\over 2 T^2}}  \left[H'(t_1),\left[ H'(t_2)\right]\right] \nonumber\\
&=& -i[H^{(2)}_T,\rho]-[\mathcal{L}_T, [\mathcal{L}_T,\rho]],
\end{eqnarray}
where both
\begin{equation}\label{eq:second_energy}
H^{(2)}_T= {i\over 2\sqrt\pi T}\int_{-\infty}^\infty dt_1
\int_{-\infty}^{t_1} dt_2 \; e^{-{t_1^2+t_2^2\over 2 T^2}}
[H'(t_1),H'(t_2)]
\end{equation}
and
\begin{equation}\label{eq:lindop}
\mathcal{L}_T=\sqrt{1\over 2\sqrt\pi T} \int_{-\infty}^\infty dt
\; e^{-{t^2\over 2 T^2}}  H'(t)
\end{equation}
are self-adjoint operators on $\mathcal{H}$. This puts
$\widetilde{K}_T$ in explicit Lindblad form, and shows that
$Z+\lambda^2 \widetilde{K}_T$ is a generator of a (completely
positive) quantum dynamical semigroup (QDS) for any finite value
of $T$ (see \cite{lindblad}). This is possible thanks to the
symmetric definition of $K_R$: a similar treatment of $K_D$
wouldn't allow to reach these conclusions, and this explains the
effort we've made so far to justify the semigroup generated by
$K_R$ in the weak-limit. Of course we're not yet completely done
as far as positivity is concerned: one would like to show
positivity for the projected $K_T$ rather than for the unphysical
$\widetilde{K}_T$ (we'll comment on this later). We shall see in
the sequel that under fairly general conditions on $P_0$, $K_T$
generates a completely positive QDS whenever $\widetilde{K}_T$
does.

Another key issue concerns the weak-limit: if $T$ is kept fixed,
than of course the generated semigroup fails to describe the exact
evolution when $\lambda\rightarrow 0$, while all of our
approximations hold only in that case. The idea is then to allow a
$\lambda$-dependence
\begin{equation}\label{eq:Tscaling}
T(\lambda)\sim |\lambda|^{-\xi} \widetilde{T}, \quad
\lambda\rightarrow 0
\end{equation}
(we shall soon be more precise on this asymptotic behavior, thanks
to our next theorem). This dispenses us to actually compute the
limit $\lim_{T\rightarrow \infty} K_T$, \emph{that would not be
defined in general}, before performing the weak-limit
$\lambda\rightarrow 0$, and simply merges the weak-coupling and
the collision limits together. The physical origin of $T$ again
helps us to understand content of this rather abstract procedure:
when we decrease $\lambda$ we have to increase the time variable
with $\lambda^{-2}t$, and so also the time $T$ needed to observe
the system, and resolve its frequencies, increases. From the
mathematical point of view, constructing the averaging map
together as the weak-limit $\lambda\rightarrow 0$ is performed,
allows $K_T=K_{T(\lambda)}$ to be well defined, and to generate a
completely positive QDS, every step of the way, and as we now
state, to correctly describe the exact dynamics in the weak-limit.

Now we recall that we defined as
\begin{equation}
\widetilde{W}^\lambda_t=e^{(Z_0+\lambda^2 K_R)t}
\end{equation}
the semigroup dynamics generated by $K_R$ and we give the
following

\begin{theorem}\label{thr:K_T}
Suppose an "observation time" $T(\lambda)$ is given, with the
asymptotic behavior reported in (\ref{eq:Tscaling}) for some
$\widetilde{T}>0$ and real $\xi$, and call
\begin{equation}
\widehat{W}^\lambda_t=e^{(Z_0+\lambda^2 K_{T(\lambda)})t}
\end{equation}
the generated dynamics. Suppose that
\begin{equation}
\int_0^\infty \|A_{01}U_t A_{10}\| \, dt < \infty.
\end{equation}

Then for every $\overline{\tau}>0$, and $0<\xi<2$ we have
\begin{equation}
\lim_{\lambda\rightarrow 0} \left\{ \sup_{0\leq t\leq
\lambda^{-2}\overline{\tau}}
\left\|\widetilde{W}^\lambda_t-\widehat{W}^\lambda_t\right\|\right\}=0.
\end{equation}
\end{theorem}
As with our previous theorem, we defer the proof to Appendix B for
the mathematically inclined reader. The physical argument that
lets us understand why we must suppose the asymptotics $0<\xi<2$
refers to some difficulty already encountered: here we are saying
that \emph{not only} the subsystem time-rescaled, interaction
picture evolution, moves slowly with respect to the $r$-variable
in (\ref{eq:rossi_approx}), but \emph{it also moves slowly with
respect to our observation time} $T(\lambda)$. This obviously
fixes the scale $\xi<2$. The remaining inequality $0<\xi$ has
already been discussed when we underlined the need to make $T$ go
to infinity in the weak-coupling.

As promised, we can now make some physical guess on the
observation time $T$: we know that it must scale with
$|\lambda|^{-\xi}$ with $0<\xi<2$ strictly. The most natural case
$\xi=1$ seems appropriate. Now, the meaning of $T$ is that of the
time needed to observe the subsystem: we could argue that we must
observe it for a long time, if the interaction $A$ is small. So we
may tentatively propose
\begin{equation}\label{eq:collision}
T(\lambda)={1\over |\lambda| \, \|A\|}
\end{equation}
In fact, this is to us the only "physical" time that can be
constructed to satisfy the stated properties, starting from known
features belonging to the original global system dynamics: in
fact, it is clear that $T$ must be ascribed to some relaxation
process, which, in our theory, are brought by the perturbation $A$
---the energy brought by $Z$, instead, just makes the system (and
the subsystem) rotate with unitarities---. Here the norm of the
perturbation $A$ has a simple meaning: it is the greatest
eigenvalue of $A$. This is indeed finite, as we are supposing $A$
to be bounded, and represents the greatest internal energy
transition, brought by the perturbation, that leaves some density
matrix invariant. Then, the highest is such a transition, the
smaller the amount of time we need to observe a subsystem
relaxation. This explains the name "completed collision time":
according to Eq. (\ref{eq:collision}), the more energy is brought
by our perturbation, the less is the time required to observe the
subsystem relaxation. This is precisely what happens with our
radio: the time we need to reach some fixed tuning precision
depends on how high the frequency is, that we look for.

We have shown that the exact dynamics is well described up to
arbitrary long times $\lambda^{-2}\overline\tau>0$ by the
(markovian) semigroup $\widehat{W}^\lambda_t$ in the weak coupling
limit. Such dynamics has been shown to be always well defined,
irrespective of the subsystem spectral properties.

\section{Positivity}\label{sec:positivity}
We have already shown that in the (physically interesting and
fairly general) case $Z\rho=-i[H_0,\rho]$ for a self-adjoint $H_0$
on a Hilbert space $\mathcal{H}$ and $A\rho=-i[H',\rho]$ ($H'$
self-adjoint), then the dynamics generated by $Z+\lambda^2
\widetilde{K}_T$ is completely positive for any given $T>0$, as
the generator itself has been written in Lindblad form:
\begin{equation}
\widetilde{K}_T\rho =-i[H^{(2)}_T,\rho]-[\mathcal{L}_T,
[\mathcal{L}_T,\rho]].
\end{equation}
\emph{This generator is physically completely meaningless if not
projected, as neither it describes the global, hamiltonian,
dynamics, nor it describes the subsystem projected dynamics}: we
shall make use of it only for convenience in some intermediate
step, always remembering that the final results only come from its
projected version $K_T$.

We now proceed to a simple analysis to determine some fairly
general conditions on the projection $P_0$ under which the
superoperator $\mathbb{L}_{\lambda T}=Z_0+\lambda^2 K_T$ can also
be put in Lindblad form, and thus generate a (completely positive)
QDS. In what follows, we shall make use of the dual
$\widetilde{P}_0$ of $P_0$, written in Heisenberg picture: the two
objects will thus be linked by the duality
$\mbox{Tr}(P_0(\rho)X)=\mbox{Tr}(\rho \widetilde{P}_0(X))$.

\begin{theorem}\label{thr:positivity}
Suppose $\widetilde{P}_0$ is a completely positive projection on
the observable space $B(\mathcal{H})$, so that it has Krauss
decomposition
\begin{equation}
\widetilde{P}_0 X=\sum_{\alpha\in\mathcal{I}} V_\alpha^\dagger X
V_\alpha
\end{equation}
for some operators $V_\alpha$ on $\mathcal{H}$, and (possibly
uncountably infinite) indexing set $\mathcal{I}$.

Suppose that $\widetilde{P}_0$ maps the observable space
$B(\mathcal{H})$ onto the subalgebra of observables $\mathcal{X}$
defined as follows: $X\in\mathcal{X}$ if and only if for every
$\alpha$,
\begin{equation}
[X,V_\alpha]=[X,V_\alpha^\dagger]=0 \;.
\end{equation}

Suppose also that $\widetilde{P}_0(1)=1$, and that
$\widetilde{P}_0$ is dual to the (completely positive projection)
$P_0$ on the state space of trace-class operators
$\mathcal{T}(\mathcal{H})$.

Then for all real $\lambda$ and $T>0$, the operator
$\mathbb{L}_{\lambda T}=Z_0+\lambda^2 K_T$ generates a (completely
positive and trace preserving) Quantum Dynamical Semigroup.
\end{theorem}

\begin{proof}
We consider
\begin{equation}\label{eq:projected_supop}
\mathbb{L}_{\lambda T}=P_0 (Z+\lambda^2\widetilde{K}_T) P_0,
\end{equation}
and  note that $Z+\lambda^2\widetilde{K}_T$ is dual to a generator
\begin{equation}
\widetilde{L}_{\lambda T} X=i[\widetilde{H}_{\lambda
T},X]-{\lambda^2 \over 2} \{\mathcal{L}_T \mathcal{L}_T, X\} +
\lambda^2 \mathcal{L}_T X \mathcal{L}_T
\end{equation}
of a completely positive QDS on $B(\mathcal{H})$, for the
self-adjoint
\begin{equation}
\widetilde{H}_{\lambda T}=H_0+\lambda^2 H_T^{(2)}
\end{equation}
(see (\ref{eq:second_energy}) and (\ref{eq:lindop}) for explicit
expressions). Then for $X\in\mathcal{X}$ and
$\rho\in\mathcal{B}_0$ we have
\begin{eqnarray}
\textit{Tr}\left(\mathbb{L}_{\lambda T}(\rho) X\right) &=& \textit{Tr}\left(\rho \widetilde{P}_0\{(Z+\lambda^2 \widetilde{K}_T) X\}\right) \nonumber \\
&=& \textit{Tr}\left(\rho \widetilde{\mathbb{L}}_{\lambda
T}(X)\right) \;,
\end{eqnarray}
for the superoperator
\begin{equation}
\widetilde{\mathbb{L}}_{\lambda T}=\widetilde{P}_0 (Z+\lambda^2
\widetilde{K}_T)
\end{equation}
on $\mathcal{X}$. Using the fact that every $X\in\mathcal{X}$
commutes with $V_\beta^{(\dagger)}$ we compute
\begin{eqnarray}
\widetilde{\mathbb{L}}_{\lambda T} X &=& i[\sum_\beta V_\beta^\dagger\widetilde{H}_{\lambda T} V_\beta,X] \\
&-&{\lambda^2 \over 2}\sum_{\beta} \{V_\beta^\dagger \mathcal{L}_T
\mathcal{L}_T V_\beta, X\} + \lambda^2 \sum V_\beta^\dagger
\mathcal{L}_T X \mathcal{L}_T V_\beta \nonumber
\end{eqnarray}
Inserting the completeness relation $\sum_\alpha V_\alpha^\dagger
V_\alpha=1$, which follows from $\widetilde{P}_0(1)=1$, we obtain
the Lindblad form
\begin{widetext}
\begin{eqnarray}
\widetilde{\mathbb{L}}_{\lambda T} X&=&i[\sum_\beta V_\beta^\dagger\widetilde{H}_{\lambda T} V_\beta,X]-{\lambda^2 \over 2}\sum_{\alpha,\beta} \{(V_\beta^\dagger \mathcal{L}_T V_\alpha^\dagger)( V_\alpha \mathcal{L}_T V_\beta), X\} + \lambda^2 \sum_{\alpha,\beta} (V_\beta^\dagger \mathcal{L}_T V_\alpha^\dagger) X (V_\alpha \mathcal{L}_T V_\beta) \nonumber\\
&=& i[\widetilde{P}_0\left(H_0\right),X]+
\lambda^2\left(i\left[\widetilde{P}_0\left({H}_{T}^{(2)}\right),X\right]-{1\over
2}\sum_{\alpha\beta} \{{D_{\alpha\beta}^T}^\dagger
D_{\alpha\beta}^T,X\} + \sum_{\alpha\beta}
{D_{\alpha\beta}^T}^\dagger X D_{\alpha\beta}^T \right)
\end{eqnarray}
\end{widetext}
for scattering operators
\begin{equation}\label{eq:dalphabeta}
D_{\alpha\beta}^T=V_\alpha \mathcal{L}_T V_\beta
\end{equation}
defined on $\mathcal{X}$. This shows that $\mathbb{L}_{\lambda
T}=Z_0+\lambda^2 K_T$ generates a completely positive QDS on
$\mathcal{B}_0$ through its dual $\widetilde{\mathbb{L}}_{\lambda
T}$ on $\mathcal{X}$: this is so indeed, precisely because
\begin{equation}
\mbox{Tr}\left(e^{\mathbb{L}_{\lambda T} t}(\rho) \, X\right) =
\mbox{Tr}\left(\rho \, e^{\widetilde{\mathbb{L}}_{\lambda T}
t}(X)\right) \;.
\end{equation}
So this completes the proof.
\end{proof}
As we see, the proof is fully constructive and gives an explicit
form for our Lindblad generator, starting directly from
fundamental objects that define the model, either algebraically
through the projection, or dynamically through the global system
hamiltonian.

Before making some examples to show the flexibility of this whole
theory, let us comment here on what we believe to be a widely
spread misunderstanding: the projected superoperator
$Z+\lambda^2\widetilde{K}_T$, for example in
(\ref{eq:projected_supop}), cannot be chosen at will. In other
words, to take a general Lindblad operator $\mathbb{L}$ and
project it with some given $P_0$, for studying the positivity of
the dynamics $e^{P_0\mathbb{L}\:t}$, is not meaningful. We have
shown that it \emph{is} meaningful when $\mathbb{L}$ comes from
some markovian approximation \emph{that already involves the
projection} $P_0$, through the condition that no subsystems first
order energy renormalisation is present ($A_{00}=P_0 A P_0 =0$).
It could be argued that this last condition is not at all
necessary, and only comes from simplicity in the proofs involved.
This is true indeed, but it does not gauge the problem away: in
the general case $A_{00}\neq 0$, the Nakajima, Prigogine,
Resibois, and Zwanzig master equation (\ref{eq:exact}) tells us
that, whatever the markovian approximation may be, it should have
a projection $P_0$ to the right as well as to the left, and an
unavoidable $P_1=1-P_0$ in the middle. So once again, \emph{this
rich and intrinsic dependence between the dynamics and the
projection cannot be causally factored}: no one comes first. Said
in other words, the (highly nontrivial) dynamics of interest $P_0
e^{\mathbb{L}\:t} P_0$, is simply \emph{different from}
$e^{P_0\mathbb{L}P_0\:t}$, at least in general. We restate that,
if $A_{00}=0$ and $[Z,P_0]=0$, we have proven that
\begin{equation}
P_0 e^{(Z+\lambda A)\:t} P_0\;\sim\; e^{(Z_0+\lambda^2
K_{T(\lambda)})t}, \quad \lambda\sim 0
\end{equation}
up to $\lambda^{-2}$-rescaled times (we refer to our Theorem
\ref{thr:K_T} for a more precise formulation), and the right hand
side is a positive map at any time if $P_0$ fulfills the
hypotheses of our Theorem \ref{thr:positivity}.

We can make a trivial counterexample along these lines: if one is
given a global hamiltonian dynamics $\mathbb{L}\rho=-i[H_0+\lambda
H',\rho]$, and $\rho$ is in the image of $P_0$, then
$P_0\mathbb{L}\rho=-i[P_0(H),\rho]$ not only doesn't give any
dissipative effect, but also (we have now abundantly shown that)
\emph{it doesn't catch up with the exact projected dynamics} $P_0
e^{\mathbb{L}t} P_0$ (that we have been studying all throughout),
not even in the case of weak coupling limit when $H=H_0+\lambda^2
H'$ and $\lambda$ is small, and not even in case $[Z,P_0]=0$.

This shows that the different projection techniques, that have
been studied so far by various groups, all suffered from the
\emph{dynamical} restrictions present in Davies' theory
\cite{davies1}. That is, one was forced to require (a global
hamiltonian dynamcs and) that the subsystem hamiltonian has
discrete spectrum. Our work here has thus been devoted to free
oneself from this restriction. This allows one to study
projections (i.e. subsystems) in a much more general contest, and
we shall give an important example of this, straight away. We have
to say that our analysis still suffers from the restriction
$A_{00}=0$. But so did Davies's analysis in \cite{davies1}, and
moreover we are currently obtaining already very encouraging
results in the more general case $A_{00}\neq 0$.
\subsection{Quantum Fermi's Golden Rule}
As a first important example, suppose $V_\alpha=V_\alpha^\dagger$
are mutually orthogonal projections on the Hilbert space
$\mathcal{H}$, and call the "quantum populations"
\begin{equation}
\rho_\alpha=V_\alpha \rho V_\alpha
\end{equation}
for a given $\rho\in\mathcal{B}_0$. Suppose also that $[P_0,Z]=0$,
with $Z\rho=-i[H_0,\rho]$, so that $H_0=\sum_\alpha H_\alpha$ with
\begin{equation}
V_\beta H_\alpha
V_{\beta'}=\delta_{\alpha\beta}\delta_{\alpha\beta'}H_\alpha \;,
\end{equation}
and put $A\rho=-i[H',\rho]$. Then the operators
$D_{\alpha\beta}^T$ in (\ref{eq:dalphabeta}) can be interpreted as
"quantum transition rates" among the "quantum populations"
$\{\rho_\alpha\}$, and one obtains an evolution
\begin{widetext}
\begin{equation}\label{eq:qfgr}
\partial \rho_\alpha = -i[H_\alpha,\rho_\alpha] +\lambda^2\left( -i [H_{T,\alpha}^{(2)},\rho_\alpha] -{1\over 2}\sum_\beta \{{D_{\beta\alpha}^T}^\dagger D_{\beta\alpha}^T,\rho_\alpha\} + \sum_\beta D_{\alpha\beta}^T \rho_\beta {D_{\alpha\beta}^T}^\dagger \right)
\end{equation}
\end{widetext}
(with $H_{T,\alpha}^{(2)}=V_\alpha H_{T}^{(2)} V_\alpha$) that in
fact couples the different "quantum populations"/density matrices
and guarantees positivity of each $\rho_\alpha$, as one can easily
see. Since the projections are mutually orthogonal ($V_\alpha
V_\beta=\delta_{\alpha\beta}V_\beta$), the sum over the index
$\beta$ in last equation can be restricted on all values such that
$\beta \neq \alpha$, as the condition $A_{00}=0$ amounts to
$D_{\alpha\alpha}^T=0$, as can easily be seen.

The generator in eq.(\ref{eq:qfgr}) for $P_0\rho$, or
alternatively this coupled linear system for density matrices
$\rho_\alpha$, is of special interest, as  it is a clear
generalization of the classical Fokker-Planck equation, and on the
other hand it is peculiar to the case $\mathcal{B}_0$ is not
finite dimensional and $Z_0$ has continuous spectrum. In fact,
only in these conditions one can hope to show that
\begin{equation}
\int_0^\infty \|A_{01}U_t A_{10}\| \, dt < \infty,
\end{equation}
which is a necessary requirement for all the theorems on the weak
limit that have been proven by us here and by Davies
in~\cite{davies1,davies2}. Again, as said before, one could say
that only when the free hamiltonian spectrum is continuous, the
"polarizations" $V_\alpha\rho V_\beta$ ($\alpha \neq \beta$)
contain enough degrees of freedom to allow an exponentially
decaying solution instead of Bloch oscillations. Current work on
the possibility of applying these ideas to the theory of quantum
transport are being addressed by us. As it is, equation
(\ref{eq:qfgr}) could thus be addressed as a "Quantum Fermi's
Golden Rule", as the "quantum populations" $\rho_\alpha$ are in
fact (positive definite) density matrices rather then (positive)
real numbers.

Note also that Eq.(\ref{eq:qfgr}) is a linear homogeneous system
of (dissipative) master equations that guarantees conservation of
a "global trace", i.e. $\mbox{Tr}(\sum_\alpha
\rho_\alpha)=\mbox{const}$, and positivity at all times for each
quantum population $\rho_\alpha$. This presents striking formal
similarities with a very recent work \cite{breuerpra}, where the
author discusses a most interesting "entangling" projection on
bipartite systems that could be thought of as a generalization of
the usual partial trace projection. There, as here, the author
arrives to a linear homogeneous system of (dissipative) master
equations describing (in his case) the dissipative/decoherent
interplay among the different entangled sectors that make up the
subsystem density matrix. Now, from one side, we will show in a
moment that this "entangling" projection can be easily
incorporated in our theory, as it is a particular (but maybe not
so much) case of the projections we consider in
Theorem~\ref{thr:positivity}. From another side, our example here
is different in that \emph{it doesn't refer to a bipartite
system}: our subsystem is, so to say, closed, with no environment
whatsoever, and with none of its subsectors $\rho_\alpha$
decohering to any given thermal state. This could implement the
idea of many systems interacting together, and finally decohering
to a steady state just because information continuously (and
irreversibly) flows from them to their quantum polarization space.
This we feel is much closer to the original idea that led Fermi to
his celebrated Fermi's Golden Rule~\cite{FGR}, as contrary to what
has been done in recent years (see for example~\cite{open}), he
did \emph{not} consider bipartite systems, nor the idea that
information irreversibly flew from a system to an environment.
Rather, information was lost in the system polarizations for some
unperturbed hamiltonian basis, and this is exactly what we do
here, extending his Fokker-Planck dynamics~\cite{fokker} among
(positive) populations to a Quantum Fokker-Planck dynamics among
(positive definite) density matrices.

It could be argued that our generator is unphysical, or
unsubstantial, as it completely disappears (it goes to zero) in
the limit $T\rightarrow 0$. We hope to have here reported enough
physical (and mathematical) evidence that such a procedure
\emph{should not} be undertaken (this is clear for example by
looking at Eq.~(\ref{eq:collision})). Rather, one could, for
example, study the steady states $\rho=\rho(T)$ as a function of
the collision time $T$, and \emph{only at that point} perform the
limit $\rho_{\mbox{steady}}=\lim_{T\rightarrow \infty} \rho(T)$.
This limit will of course give a non-trivial result. The situation
here is very similar to that of the weak coupling limit, at
variance with the thermodynamical limit: the two limits simply
\emph{do not commute}, and the order to be taken depends upon
one's interest, as justified on physical grounds. Our argument for
the completed collision limit $T\rightarrow \infty$ is, again,
that the semigroup approximation holds only for intermediate times
$T$, between the subsystem relaxation times and the global
Poincar\'e recurrences (see for example~\cite{FGR}, or the
brilliant exposition in~\cite{cohen}). Thus, it must be finite
every step of the way, and be brought to infinity only in a second
step. There is one notable exception to this: precisely when the
free hamiltonian spectrum is discrete, there will be a finite $T$
(as said before) over which we'll clearly distinguish among all
the different frequencies. If we used a higher $T'>T$, the physics
will then be left unchanged, exactly because we already reached
infinite precision with $T$. This explains, physically and
mathematically, the reason why one could free himself of $T$ in
the discrete case, by letting it go to infinity.

So we understand why we all had to study the FGR in the (rather
cumbersome) case of a free hamiltonian with mixed
spectrum~\cite{cohen}: on one side, one needs a continuous
spectrum to hope to see dissipation, but on the other side, the
diagonal part of the scattering operator is well defined only in
the discrete case. The only way out seems to force the interacting
hamiltonian communicate between the discrete and the continuous
part of the free one! Here what we do is essentially to relax the
condition that the relevant degrees of freedom are the free
hamiltonian eigenvalues, and rather take them as diagonal boxes.
Our model is thus fully preserving the quantum features within
each sector.

\subsection{Entangling Projection}
As promised, we shall now make contact with the work in
\cite{breuerpra} and show how the projections there considered can
be implemented in our formalism. We just here anticipate that this
is going to be a generalization of the standard trace-projection
over the bath. Let $\mathcal{H}=\mathcal{H}_A\otimes\mathcal{H}_B$
be the tensor product of two Hilbert spaces, representing a
coupled system-environment pair. Let
${\mathcal{B}}=\mathcal{T}(\mathcal{H})$ be the space of
trace-class operators on $\mathcal{H}$, and $B(\mathcal{H})$ the
observable algebra of bounded operators on $\mathcal{H}$ (the same
goes for $\mathcal{H}_A$ and $\mathcal{H}_B$).

Take a finite number $C_n$ and $D_n$ ($n=1\ldots N$), of operators
in $B(\mathcal{H}_B)$, such that
\begin{equation}\label{hp:entproj1}
D^\dagger_n D_{n'}=\delta_{n n'} B_n
\end{equation}
for (positive definite) trace-class $B_n=D^\dagger_n D_n$ in
$\mathcal{T}(\mathcal{H}_B)$. Name the (positive definite)
operators
\begin{equation}
A_n=C_n^\dagger C_n
\end{equation}
in $B(\mathcal{H}_B)$ and suppose
\begin{equation}\label{hp:entproj2}
\sum_{n=1}^N A_n =1 \;,
\end{equation}
\begin{equation}\label{hp:entproj3}
A_n A_{n'} =\delta_{n n'} A_n \;,
\end{equation}
and
\begin{equation}\label{hp:entproj4}
\mbox{Tr}(A_n B_{n'})=\delta_{n n'} \;.
\end{equation}

Take a basis $\{|\alpha\rangle\}$ of $\mathcal{H}_B$, and define
the operators
\begin{equation}
V_{\alpha\alpha'}=1_A\otimes \sum_{n=1}^N
D_n^\dagger|\alpha\rangle\langle \alpha'| C_n
\end{equation}
on $\mathcal{H}$. With these operators define the completely
positive map $\widetilde{P}_0$ on $B(\mathcal{H})$ by
\begin{equation}\label{eq:dualproj}
\widetilde{P}_0 X=\sum_{\alpha \alpha'} V_{\alpha\alpha'}^\dagger
X V_{\alpha\alpha'} \;.
\end{equation}
This map can be checked to be a projection thanks to the
hypotheses (\ref{hp:entproj1}) and (\ref{hp:entproj4}). The
completeness hypothesis (\ref{hp:entproj2}) guarantees that
$\widetilde{P}_0(1)=1$, and hypothesis (\ref{hp:entproj3})
guarantees that $\mathcal{X}=\widetilde{P}_0(B(\mathcal{H}))$ is a
subalgebra of the global observable algebra $B(\mathcal{H})$. So
$\widetilde{P}_0$ fulfills the hypotheses of our Theorem
\ref{thr:positivity}. In particular, one can see that the
projected subalgebra is
\begin{equation}
\mathcal{X}= \{X\in B(\mathcal{H}) \; | \; X=\sum_{n=1}^N
X_n\otimes A_n \;,\; X_n\in B(\mathcal{H_A})\} \;.
\end{equation}
Using our hypotheses we then compute
\begin{eqnarray}\label{eq:proj}
P_0 \rho &=& \sum_{\alpha \alpha'} V_{\alpha\alpha'} \rho V_{\alpha\alpha'}^\dagger \nonumber \\
&=& \sum_{n} \mbox{Tr}_B(\rho(1\otimes A_n))\otimes B_n \;,
\end{eqnarray}
which is precisely Equation (12) in \cite{breuerpra}. Then it is
possible to check that $P_0$ is also a (completely positive)
projection, precisely because of  hypothesis (\ref{hp:entproj4}).
Our procedure used to obtain it is however different from
\cite{breuerpra}, and it would be surely worth the effort to
compare the two approaches further. For example, the only common
hypothesis to obtain such an entangling projection is our equation
(\ref{hp:entproj4}), which in \cite{breuerpra} is Equation (6),
while it seems to us that all other hypotheses differ in the two
approaches.

But the most fundamental difference is that here we have to
satisfy, as amply discussed before, a dynamical compatibility
hypothesis, that intrinsically links $P_0$ with the form of the
Lindblad operator that one wants to project. We have shown that if
one starts from a free hamiltonian dynamics, $Z\rho=-i[H_0,\rho]$
and perturbs with $\lambda A\rho=-i\lambda[H',\rho]$, then one
arrives to a Lindblad generator for the projected subsystem
\emph{if} $[Z,P_0]=0$, or alternatively if $[U_t,P_0]=0$ for all
times. For example, if, as normally assumed, the unperturbed
hamiltonian is of the form $H_0=H_A\otimes 1+1\otimes H_B$, this
\emph{dynamical commutation condition for the subsystem} can be
checked to be fulfilled \emph{if and only if} for every $n=1\ldots
N$ we have
\begin{equation}
[H_B,A_n]=[H_B,B_n]=0 \;.
\end{equation}
Then we have proven that \emph{if} $A_{00}=0$ our dynamics
correctly describes the exact projected dynamics in the
weak-limit, and is a (completely positive markovian) Quantum
Dynamical Semigroup. If then one would like to take $A$ as a
Lindblad generator (instead of a safe self-adjoint derivation)
things would get much more complex, as in any case one should
always prove that $X^\lambda_t=e^{(Z_0+\lambda A_{00})t}$ is a
group of isometries (see~\cite{davies2}), which gives nontrivial
conditions on $A_{00}$, when the latter is switched on (again a
condition that links the choice of $P_0$ to the dynamics). This
leaves the possibility that $A-A_{00}$ is dissipative while
$A_{00}$ is a self-adjoint derivation (i.e. it comes from a
hamiltonian): again, we are currently working on this interesting
and important path. But we have to remember that the perturbation
must have a "$\lambda$" attached to it, and the (dissipative)
subsystem will describe the projected (phenomenological and
dissipative) global dynamics, only up to second order in
$\lambda$, when $\lambda\sim 0$.

\section{Remarks and Conclusion}
Although all our results suffer from being restricted to Hilbert
spaces, we feel much confident in saying that they can be extended
to more general (and natural) Banach spaces, and we are currently
working on that promising direction.

Another severe restriction is our hypothesis $A_{00}=0$ of absence
of first order subsystem energy renormalizations. Just as before,
we currently obtaining very promising results in effort to free
the theory from this restriction.

Apart from these key remarks, we have shown that, under the only
consistency with the weak-coupling limit, the markovian
approximation of the memory terms in the quantum mechanical master
equation is far from unique. We found a generator for a Quantum
Dynamical Semigroup that generalizes more standard approaches, and
contrary to them guarantees both consistency in the weak limit and
complete positivity, irrespective of the spectral properties of
the system. We have thus been able to propose a new type of
quantum generalization of the celebrated Fermi's Golden Rule,
giving an unprecedented linear homogeneous system of (dissipative)
master equations \emph{for a closed subsystem} that guarantees
positivity at all times for each of the subsystem sectors. It is
worth noting that our model requires a continuous spectrum for the
free hamiltonian, and thus is peculiar to our general theory. We
have also shown that our theory dynamically incorporates some
recently proposed models for entangled bipartite subsystems, and
compared to them. All this opens up the way to an entirely new
formalism for modelling nowadays mesoscopic-scale electronic and
optoelectronic devices, as well as to investigations on the
fundamental, and everlasting, problem irreversibility in more
abstract quantum theories.

\begin{acknowledgements}
We would like to thank Prof. Hisao Fujita Yashima (Dept.
Mathematics, University of Turin) for having offered so many days
of invaluable help and discussions to one of the authors.
\end{acknowledgements}

\appendix
\section{Proof of Theorem (\ref{th:symmetric})}
\begin{proof}
Let $\mathcal{V}$ be the Banach space of norm continuous
$\mathcal{B}_0$-valued functions on $[0,\overline{\tau}]$, and let
$b\in\mathcal{B}_0$. Define the "interaction picture" time
rescaled solution
\begin{equation}
\overline{f}_\lambda(\tau)=U_{-\lambda^{-2}\tau}\overline{W}^\lambda_{\lambda^{-2}\tau}b.
\end{equation}
Then $\overline{f}_\lambda$ is a solution to the integral equation
\begin{equation}
\overline{f}_\lambda=b+\overline{\mathcal{H}}_\lambda
\overline{f}_\lambda,
\end{equation}
where the integral operator $\overline{\mathcal{H}}_\lambda$ is
defined by
\begin{equation}\label{eq:intop}
(\overline{\mathcal{H}}_\lambda g)(\tau)=\int_0^\tau ds
U_{-\lambda^{-2}s} K_D U_{\lambda^{-2}s} g(s).
\end{equation}
Now $\overline{\mathcal{H}}_\lambda$ is a Volterra operator, so
\begin{equation}
\|\overline{\mathcal{H}}_\lambda^n\|\leq
{c}^n\overline{\tau}^n/n!,
\end{equation}
where ${c}$ does not depend upon $\lambda$, and we have
convergence of the associated Newmann series expansion
\begin{equation}
\overline{f}_\lambda=b+\overline{\mathcal{H}}_\lambda
b+\overline{\mathcal{H}}_\lambda^2 b+\cdots
\end{equation}
So far everything can be restated for $\widetilde{f}_\lambda
\leftrightarrow \widetilde{W}^\lambda$ in exactly the same way,
leading to the corresponding definition of
$\widetilde{\mathcal{H}}_\lambda$.

Subtracting the similar expansions for $\overline{f}_\lambda$ and
$\widetilde{f}_\lambda$ we obtain
\begin{eqnarray}
&&\sup_{0\leq t\leq \lambda^{-2}\overline{\tau}} \|\overline{W}^\lambda_t b -\widetilde{W}^\lambda_t b\| = \|\overline{f}_\lambda-\widetilde{f}_\lambda\|_\infty \nonumber \\
&&\quad\leq \sum_{n=0}^\infty \| \overline{\mathcal{H}}_\lambda^n
b - \widetilde{\mathcal{H}}_\lambda^n b \|_\infty
\end{eqnarray}
It's easy to see that this series is dominated uniformly with
respect to $\lambda$, so it only remains to show that
\begin{equation}
\lim_{\lambda\rightarrow 0}
\|\overline{\mathcal{H}}_\lambda-\widetilde{\mathcal{H}}_\lambda
\| = 0.
\end{equation}
This means that for every $\epsilon>0$ there's a
$\overline{\lambda}>0$ such that $|\lambda|<\overline{\lambda}$
implies that for every $g\in\mathcal{V}$
\begin{equation}
\left\| \int_0^\tau ds\; U_{-\lambda^{-2}s} \Delta K
U_{\lambda^{-2}s} g(s) \right\|<{\epsilon\over 2^2}.
\end{equation}
Before proceeding we have to keep divercences under control: the
time variable inside the integrals that define $K_D$ and $K_R$
goes from $0$ to $\infty$: we show that it can be stopped at a
finite value. To do this, we write explicitly the definition of
the operators involved in the last inequality:
\begin{eqnarray}\label{eq:commutator}
&&\int_0^\tau ds\; U_{-\lambda^{-2}s} \Delta K U_{\lambda^{-2}s} g(s) \nonumber \\
&=& \int_0^\infty dx \int_0^\tau ds\; \left[U_{-{x\over 2}},
U_{-\lambda^{-2}s-{x\over 2}}B_x U_{\lambda^{-2}s-{x\over
2}}\right] g(s) \nonumber \\,
\end{eqnarray}
where we have put $B_x=A_{01}U_x A_{10}$. Now the integral over
the $x$-variable on $[0,\infty]$ can be split into $[0,\overline
x]\cup[\overline x,\infty]$, which gives
\begin{eqnarray}
&&\left\|\int_0^\tau ds\; U_{-\lambda^{-2}s} \Delta K U_{\lambda^{-2}s} g(s)\right\| \nonumber \\
&&\leq \left\|\int_0^{\overline x} \cdots \right\|+2 \|g\|_\infty
\overline \tau \int_{\overline{x}}^\infty \|B_x\|\;dx.
\end{eqnarray}
According to our hypothesis (\ref{eq:no_oscillation}) on $B_x$ we
can choose a suitable $\overline{x}$ such that the last term is
smaller than $\epsilon/2^3$, independently of $\lambda$:
\begin{equation}
2 \|g\|_\infty \overline \tau \int_{\overline{x}}^\infty
\|B_x\|\;dx < {\epsilon \over 2^3}.
\end{equation}
Once we fix such an $\overline{x}$, our problem becomes to find a
$0<\overline{\lambda}_2\leq\lambda_1$ such that if
$|\lambda|<\overline{\lambda}_2$, then for every
$0\leq\tau\leq\overline{\tau}$:
\begin{equation}
\left\|\int_0^{\overline{x}} dx \int_0^\tau ds\; \left[U_{-{x\over
2}}, U_{-\lambda^{-2}s-{x\over 2}}B_x U_{\lambda^{-2}s-{x\over
2}}\right] g(s)\right\| < {\epsilon \over 2^4}.
\end{equation}
We now make use of the spectral theorem: if
\begin{equation}
Z=\int_{-\infty}^\infty i\omega\; dE_\omega
\end{equation}
is the spectral decomposition of $Z$ for its spectral family
$E_\omega$ (if $Z\rho=-i[H,\rho]$ then the $\omega$'s have meaning
of energy differences, or characteristic frequencies of the global
system), then
\begin{equation}
U_t=\int_{-\infty}^\infty e^{i\omega t}\; dE_\omega.
\end{equation}
As for the time variable $x$, we split the frequency range into
$(-\infty,-\overline{\omega})
\cup[-\overline{\omega},\overline{\omega}]\cup(\overline{\omega},\infty)$
and define
\begin{equation}
U^{(\overline{\omega})}_t=\int_{-\overline{\omega}}^{\overline{\omega}}
e^{i\omega t}\; dE_\omega.
\end{equation}
Adding and subtracting
$U^{(\overline{\omega})}_{-\lambda^{-2}s-x}$ and
$U^{(\overline{\omega})}_{-\lambda^{-2}s-{x\over 2}}$ we are led
to consider the quantities
\begin{equation}\label{eq:bounded_spectra_1}
\int_0^{\overline{x}} dx \int_0^\tau ds \;
\|(U_{-\lambda^{-2}s-x}-U^{({\overline{\omega}})}_{-\lambda^{-2}s-x})
B_x U_{-\lambda^{-2}s} g(s)\|
\end{equation}
and
\begin{equation}\label{eq:bounded_spectra_2}
\int_0^{\overline{x}} dx \int_0^\tau ds
\|(U_{-\lambda^{-2}s-{x\over
2}}-U^{(\overline{\omega})}_{-\lambda^{-2}s-{x\over 2}}) B_x
U_{-\lambda^{-2}s-{x\over 2}} g(s)\|.
\end{equation}
The first is dominated by (for compactness of the involved
intervals, $\sup=\max$)
\begin{eqnarray}\label{eq:bounded_spectra_1_bis}
&&\overline{x} \overline{\tau} \; \max \left\{ \|(U_{-\lambda^{-2}s-x}-U^{({\overline{\omega}})}_{-\lambda^{-2}s-x}) B_x U_{-\lambda^{-2}s} g(s)\|, \right. \nonumber \\
&&\left. \quad 0\leq s\leq \overline{\tau}, 0\leq x\leq\overline{x} \right\}\nonumber\\
&&\leq \overline{x} \overline{\tau} \left\| \left(\int_{-\infty}^{-\overline{\omega}}+\int_{-\overline{\omega}}^{\infty}\right)  e^{i(\lambda^{-2}\underline{s}+\underline{x})\omega}\: dE_\omega\; B_{\underline{x}} U_{-\lambda^{-2}\underline{s}} g(\underline{s}) \right\| \nonumber \\
&&\leq \overline{x} \overline{\tau}
\left(\int_{-\infty}^{-\overline{\omega}}+\int_{-\overline{\omega}}^{\infty}\right)
\|dE_\omega\; B_{\underline{x}} U_{-\lambda^{-2}\underline{s}}
g(\underline{s}) \|.
\end{eqnarray}
But since the $E_\omega$'s are orthogonal projections
\begin{eqnarray}
&&\int_{-\infty}^{\infty} \|dE_\omega\; B_{\underline{x}} U_{-\lambda^{-2}{\underline{s}}} g({\underline{s}}) \| \nonumber \\
&=& \| B_{\underline{x}} U_{-\lambda^{-2}{\underline{s}}} g({\underline{s}}) \| \nonumber \\
&\leq & \|B_{\underline{x}}\| \|g({\underline{s}})\|
\end{eqnarray}
is bounded independently of $\lambda$, so we can choose
$\overline{\omega}$ such that the quantity in the last line in
(\ref{eq:bounded_spectra_1_bis}) is dominated by ${\epsilon\over
2^5}$ independently of $\lambda$ (it can however depend on $g$, as
a consequence of the estimation above, but this will not bother
us):
\begin{equation}
\overline{x} \overline{\tau}
\left(\int_{-\infty}^{-\overline{\omega}}+\int_{-\overline{\omega}}^{\infty}\right)
\|dE_\omega\; B_{\underline{x}} U_{-\lambda^{-2}\underline{s}}
g(\underline{s}) \| < {\epsilon \over 2^5}
\end{equation}
Following the same reasoning, we can choose $\overline\omega$ such
that also the difference in (\ref{eq:bounded_spectra_2}) is
controlled by ${\epsilon\over 2^6}$. In fact, again with the same
reasoning, by adding and subtracting
$U^{(\overline{\omega})}_{\lambda^{-2}s}$ and
$U^{(\overline{\omega})}_{\lambda^{-2}s-{x\over 2}}$ we easily see
that we can fix an $\overline\omega$ such that
\begin{eqnarray}
&&\int_0^{\overline{x}} dx \int_0^\tau ds \; \|U^{({\overline{\omega}})}_{-\lambda^{-2}s-x} B_x (U_{\lambda^{-2}s}-U^{({\overline{\omega}})}_{\lambda^{-2}s}) g(s)\| \nonumber \\
&&< {\epsilon\over 2^7}
\end{eqnarray}
and
\begin{eqnarray}
&&\int_0^{\overline{x}} dx \int_0^\tau ds \; \|U^{({\overline{\omega}})}_{-\lambda^{-2}s-{x\over 2}} B_x (U_{\lambda^{-2}s-{x\over 2}}-U^{({\overline{\omega}})}_{\lambda^{-2}s-{x\over 2}}) g(s)\| \nonumber \\
&&< {\epsilon\over 2^8}
\end{eqnarray}
independently of $\lambda$. So we now have to prove that, when we
fix $\overline{x}$ and $\overline{\omega}$ as done before, we can
find a $0<\overline{\lambda}_2\leq\lambda_1$ such that when
$|\lambda|<\overline{\lambda}_2$, for every
$0\leq\tau\leq\overline{\tau}$:
\begin{equation}\label{eq:allbounded}
\left\|\int_0^{\overline{x}} dx \int_0^\tau ds\;
\left[U^{({\overline{\omega}})}_{-{x\over 2}},
U^{({\overline{\omega}})}_{-\lambda^{-2}s-{x\over 2}}B_x
U^{({\overline{\omega}})}_{\lambda^{-2}s-{x\over 2}}\right]
g(s)\right\| < {\epsilon \over 2^9}.
\end{equation}
We proceed by plugging in the definition of Stiltjes integral
involved in the spectral decompositions of the unitarities $U_t$
(see for example \cite{triebel}). If
$\Sigma(-\overline\omega,\overline\omega)$ is the set of
subdivisions of the interval $[-\overline\omega,\overline\omega]$,
and $\sigma\in\Sigma(-\overline\omega,\overline\omega)$ has points
$-\overline\omega=\omega_0<\omega_1<\cdots<\omega_{N_\sigma}=\overline\omega$,
we put
\begin{equation}
d(\sigma)=\max_{k=1,\ldots,N_{\sigma}} |\omega_{k}-\omega_{k-1}|.
\end{equation}
Then the limit
\begin{equation}
\int_{-\overline{\omega}}^{\overline{\omega}} e^{i\omega t}\;
dE_\omega f =\lim_{d(\sigma)\rightarrow 0} \sum_{k=0}^{N_\sigma-1}
e^{i{\widetilde\omega}_k t} (E_{\omega_{k+1}}-E_{\omega_k})f,
\end{equation}
where $\widetilde\omega_k\in [\omega_k,\omega_{k+1}[$, exists and
defines the left hand side. Then eq. (\ref{eq:allbounded}) becomes
\begin{widetext}
\begin{eqnarray}
&&\left\| \lim_{d(\sigma_1),d(\sigma_2)\rightarrow 0} \int_0^{\overline{x}} dx \int_0^\tau ds\; \sum_{k_1=0}^{N_{\sigma_1}-1} \sum_{k_2=0}^{N_{\sigma_2}-1} \left\{ e^{i{\widetilde\omega}_{k_1}^{(1)} (-\lambda^{-2}s-x)+i{\widetilde\omega}_{k_2}^{(2)} \lambda^{-2}s} -  e^{i{\widetilde\omega}_{k_1}^{(1)} (-\lambda^{-2}s-{x\over 2})+i{\widetilde\omega}_{k_2}^{(2)} (\lambda^{-2}s-{x\over 2})}\right\} \right.\nonumber\\
&&\left. (E_{\omega_{k_1+1}^{(1)}}-E_{\omega_{k_1}^{(1)}}) B_x
(E_{\omega_{k_2+1}^{(2)}}-E_{\omega_{k_2}^{(2)}})  g(s) \right\|
\end{eqnarray}
\end{widetext}
with obvious notations. But clearly it suffices to consider only
one subdivision instead of two, and prove that there exist a
$\overline\lambda_2>0$ and a subdivision $\overline\sigma$ such
that if $|\lambda|<\overline\lambda_2$ and
$d(\sigma)<d(\overline\sigma)$ then
\begin{widetext}
\begin{eqnarray}
&&\left\| \int_0^{\overline{x}} dx \int_0^\tau ds\; \sum_{k_1,k_2=0}^{N_{\sigma}-1}\left\{ e^{i{\widetilde\omega}_{k_1} (-\lambda^{-2}s-x)+i{\widetilde\omega}_{k_2} \lambda^{-2}s} -  e^{i{\widetilde\omega}_{k_1} (-\lambda^{-2}s-{x\over 2})+i{\widetilde\omega}_{k_2} (\lambda^{-2}s-{x\over 2})}\right\}\right. \nonumber\\
&&\left.(E_{\omega_{k_1+1}}-E_{\omega_{k_1}}) B_x
(E_{\omega_{k_2+1}}-E_{\omega_{k_2}})  g(s) \right\|<
{\epsilon\over 2^9},
\end{eqnarray}
\end{widetext}
that is, rearranging terms,
\begin{eqnarray}
&&\left\| \sum_{k_1,k_2=0}^{N_{\sigma}-1} \int_0^{\overline{x}} dx \; e^{-i{x\over 2} {\widetilde\omega}_{k_1}} \left\{e^{-i{x\over 2}{\widetilde\omega}_{k_1}} -e^{-i{x\over 2}{\widetilde\omega}_{k_2}} \right\} \right.\nonumber\\
&&(E_{\omega_{k_1+1}}-E_{\omega_{k_1}}) B_x (E_{\omega_{k_2+1}}-E_{\omega_{k_2}}) \nonumber\\
&&\left.\int_0^\tau ds\; e^{-i \lambda^{-2}s
({\widetilde\omega}_{k_1}-{\widetilde\omega}_{k_2})} g(s)
\right\|< {\epsilon\over 2^9}.
\end{eqnarray}
But the left hand side is dominated by
\begin{eqnarray}\label{eq:analysis}
&& \sum_{k_1,k_2=0}^{N_{\sigma}-1} \int_0^{\overline{x}} dx \; \left|e^{-i{x\over 2}{\widetilde\omega}_{k_1}} -e^{-i{x\over 2}{\widetilde\omega}_{k_2}} \right| \nonumber \\
&&\left\|(E_{\omega_{k_1+1}}-E_{\omega_{k_1}}) B_x (E_{\omega_{k_2+1}}-E_{\omega_{k_2}}) \right.\nonumber\\
&&\left.\int_0^\tau ds\; e^{-i \lambda^{-2}s
({\widetilde\omega}_{k_1}-{\widetilde\omega}_{k_2})} g(s)
\right\|.
\end{eqnarray}
and we can safely control the slowly varying phase oscillations by
\begin{equation}\label{eq:slowdiag}
\left|e^{-i{x\over 2}{\widetilde\omega}_{k_1}} -e^{-i{x\over
2}{\widetilde\omega}_{k_2}} \right| \leq
|{\widetilde\omega}_{k_1}-{\widetilde\omega}_{k_2}| \;
{\overline{x} \over 2}
\end{equation}
(this explains the splitting in the time variable $x$). In
passing, we note that we can discard the contribution with equal
frequencies $k_1=k_2$ in the sum for the subdivision $\sigma$.

As the rapidly oscillating integral is concerned, we take a
$\lambda$-smoothed version of the Dirac delta functional: for a
suitable real number $\zeta>0$ we define
\begin{equation}
\delta_\lambda(x)={1\over \pi}{\sin \left({x\over
|\lambda|^{\zeta} \overline\tau}\right) \over  x}={1\over 2\pi
}\int_{-\Omega_\lambda}^{\Omega_\lambda} d\Omega\; e^{i \Omega x},
\end{equation}
with $\Omega_\lambda=|\lambda|^{-\zeta} / \overline\tau$, and
consider the approximation
\begin{widetext}
\begin{eqnarray}
&& \int_0^\tau ds\; e^{-i \lambda^{-2}s ({\widetilde\omega}_{k_1}-{\widetilde\omega}_{k_2})} g(s) \sim  \int_0^\tau ds\; e^{-i \lambda^{-2}s ({\widetilde\omega}_{k_1}-{\widetilde\omega}_{k_2})} \int_0^\tau dr\; g(r) \delta_\lambda(s-r) \nonumber \\
&=& {1\over 2\pi }\int_{-\Omega_\lambda}^{\Omega_\lambda}
d\Omega\; \int_0^\tau ds\; e^{-i \lambda^{-2}s
({\widetilde\omega}_{k_1}-{\widetilde\omega}_{k_2})-i\Omega s}
\int_0^\tau dr\; e^{i\Omega r} g(r)
\end{eqnarray}
valid for $\lambda \sim 0$. Then the term in
eq.(\ref{eq:analysis}) is bounded by
\begin{equation}\label{eq:analysis2}
\sum_{k_1,k_2=0}^{N_{\sigma}-1} \int_0^{\overline{x}} dx \;
|{\widetilde\omega}_{k_1}-{\widetilde\omega}_{k_2}| \;
{\overline{x} \over 2} {1\over \pi
}\int_{-\Omega_\lambda}^{\Omega_\lambda} d\Omega\; {1\over
|\lambda^{-2}
({\widetilde\omega}_{k_1}-{\widetilde\omega}_{k_2})+\Omega|}
\int_0^\tau dr\; \left\|(E_{\omega_{k_1+1}}-E_{\omega_{k_1}}) B_x
(E_{\omega_{k_2+1}}-E_{\omega_{k_2}})  g(r) \right\|.
\end{equation}
\end{widetext}
Now the integral in $d\Omega$ can be computed easily if, for
sufficiently small values for $\lambda$, the integrand is never
singular. This is the case provided that $\zeta < 2$, as we shall
assume. Then the integrand is either always positive or always
negative, so
\begin{equation}
\int_{-\Omega_\lambda}^{\Omega_\lambda} d\Omega\; {1\over
|\lambda^{-2}
({\widetilde\omega}_{k_1}-{\widetilde\omega}_{k_2})+\Omega|} \sim
{\lambda^{2-\zeta} \over
|{\widetilde\omega}_{k_1}-{\widetilde\omega}_{k_2}|
\overline\tau}.
\end{equation}
Putting things together, we estimate eq.(\ref{eq:analysis}) by
\begin{eqnarray}
&& \sum_{k_1,k_2=0}^{N_{\sigma}-1}  |{\widetilde\omega}_{k_1}-{\widetilde\omega}_{k_2}| \; {\overline{x} \over 2} {1\over \pi } {\lambda^{2-\zeta} \over ({\widetilde\omega}_{k_1}-{\widetilde\omega}_{k_2}) \overline\tau}  \int_0^{\overline{x}} dx \; \nonumber \\
&& \int_0^\tau dr\; \left\|(E_{\omega_{k_1+1}}-E_{\omega_{k_1}}) B_x (E_{\omega_{k_2+1}}-E_{\omega_{k_2}})  g(r) \right\| \nonumber \\
&=& {\overline{x} \over 2} {1\over \pi } {\lambda^{2-\zeta} \over
\overline\tau}  \int_0^{\overline{x}} dx \;
\int_0^\tau dr\; \left\| B_x  g(r) \right\| \nonumber \\
&\leq& { \lambda^{2-\zeta}\, \overline{x}\over 2 \pi } \;
\|g\|_\infty \int_0^{\infty} dx \|B_x\|
\end{eqnarray}
(the second line follows because the $E_\omega$'s are orthogonal
projections). Now this goes to zero as $\lambda\rightarrow 0$,
thus completing the proof.
\end{proof}
\section{Proof of Theorem (\ref{thr:K_T})}
\begin{proof}
(Throughout when there's no possible misunderstanding we should
use the notation $K_T$ instead of the complete $K_{T(\lambda)}$).
With a completely analogous notation, we follows the very same
steps in the proof of theorem \ref{th:symmetric}: we then have to
show that
\begin{equation}
\lim_{\lambda\rightarrow 0}
\|\widetilde{\mathcal{H}}_\lambda-\widehat{\mathcal{H}}_\lambda \|
= 0,
\end{equation}
where both $\widetilde{\mathcal{H}}_\lambda$ and
$\widehat{\mathcal{H}}_\lambda$ are defined analogously to
eq.(\ref{eq:intop}), respectively for $K_R$ and $K_{T(\lambda)}$.
Now defining $\stackrel{\vartriangle}{\mathcal{H}}_\lambda$ by
\begin{eqnarray}
&&(\stackrel{\vartriangle}{\mathcal{H}}_\lambda g)(\tau)=\int_0^\tau ds\; U_{-\lambda^{-2}s} \nonumber \\
&& \times\left\{{1\over \sqrt\pi T} \int_{-\infty}^\infty dq\; e^{-q^2/T^2} U_{-q} K_R U_q\right\}  U_{\lambda^{-2}s} g(s), \nonumber \\
\end{eqnarray}
where $g\in\mathcal{V}$, leads to the estimation
\begin{equation}
\|\widetilde{\mathcal{H}}_\lambda-\widehat{\mathcal{H}}_\lambda \|
\leq
\|\widetilde{\mathcal{H}}_\lambda-\stackrel{\vartriangle}{\mathcal{H}}_\lambda
\|+
\|\stackrel{\vartriangle}{\mathcal{H}}_\lambda-\widehat{\mathcal{H}}_\lambda
\|.
\end{equation}
The easiest part to estimate is the last term: one can see that
\begin{eqnarray}\label{eq:1Test}
&&\!\!\!\!\!\!\!\!\!\!\!\!\!\!\!\!\|(\stackrel{\vartriangle}{\mathcal{H}}_\lambda g)(\tau) - (\widehat{\mathcal{H}}_\lambda g)(\tau)\|\nonumber\\
&&\!\!\!\!\!\!\!\!\!\!\!\!\!\!\!\!\leq \overline{\tau} \:\|g\|_\infty \: \left\|K_R-K^{T(\lambda)}\right\| \nonumber\\
&&\!\!\!\!\!\!\!\!\!\!\!\!\!\!\!\!\leq \overline{\tau}
\:\|g\|_\infty \!\! \int_0^\infty\!\! dx (1-e^{-({x\over
2})^2/T^2(\lambda)}) \|A_{01} U_x A_{10} \|.
\end{eqnarray}
Now for any given $x\in[0,\infty)$ one has
\begin{equation}
\lim_{\lambda\rightarrow 0} (1-e^{-({x\over 2})^2/T^2(\lambda)})
=0
\end{equation}
because of the hypothesis $T(\lambda)\sim |\lambda|^{-\xi}
\widetilde{T}$ with $\xi>0$. This together with the boundedness
hypothesis
\begin{equation}
\int_0^\infty dx\; \|A_{01} U_x A_{10} \| <\infty
\end{equation}
implies that the last term in (\ref{eq:1Test}) goes to zero as
$\lambda\rightarrow 0$.

It remains to show that
\begin{equation}
\|\widetilde{\mathcal{H}}_\lambda-\stackrel{\vartriangle}{\mathcal{H}}_\lambda
\| \rightarrow 0, \quad \lambda \rightarrow 0,
\end{equation}
or, more explicitly, we have to show convergence to zero, for any
$g\in\mathcal{V}$ and uniformly on $\tau$, of
\begin{eqnarray}\label{eq:2Test}
&&\left\| \int_0^\tau ds\; U_{-\lambda^{-2}s} \left\{{1\over \sqrt\pi T} \int_{-\infty}^\infty dq\; e^{q^2/T^2} U_{-q} K_R U_q -K_R \right\} \right.\nonumber\\
&&\times\left. U_{\lambda^{-2}s} g(s) \right\|.
\end{eqnarray}
Now the very same arguments we used for theorem
(\ref{th:symmetric}) apply here, so  $U_{\pm q}$ can be
substituted with $\widetilde{U}_{\pm q}$, for a suitable cut-off
frequency $\widetilde\omega$: the difference with
eq.(\ref{eq:2Test}) is shown to be arbitrarily small in exactly
the same way as before. Again as before, we can evaluate the limit
in the Stiltjes integral for the spectral decomposition over a
single common spectral subdivision
$\sigma=\{-\widetilde\omega=\omega_0<\ldots<\omega_{N_\sigma}=\widetilde\omega\}$.
So proceeding as in theorem (\ref{th:symmetric}) we are now left
with the problem of finding, for any given $\epsilon>0$, a
suitable $\overline\lambda>0$ and spectral subdivision
$\overline\sigma$ such that for any $|\lambda|<\overline\lambda$
and $d(\sigma)<d(\overline\sigma)$ we have
\begin{eqnarray}
&&\left\| \sum_{k_1,k_2=0}^{N_{\sigma}-1} \left\{\left({1\over \sqrt\pi T} \int_{-\infty}^\infty dq\; e^{-q^2/T^2} e^{iq({\widetilde\omega}_{k_1}-{\widetilde\omega}_{k_2})}\right)-1\right\} \right.\nonumber\\
&&\times \left.(E_{\omega_{k_1+1}}-E_{\omega_{k_1}}) K_R (E_{\omega_{k_2+1}}-E_{\omega_{k_2}}) \right.\nonumber\\
&&\times\left. \int_0^\tau ds\; e^{-i \lambda^{-2}s ({\widetilde\omega}_{k_1}-{\widetilde\omega}_{k_2})} g(s) \right\| \nonumber \\
&\leq& \sum_{k_1,k_2=0}^{N_{\sigma}-1} \left| e^{-{T^2({\widetilde\omega}_{k_1}-{\widetilde\omega}_{k_2})^2}}-1\right| \nonumber\\
&&\times\left\|(E_{\omega_{k_1+1}}-E_{\omega_{k_1}}) K_R (E_{\omega_{k_2+1}}-E_{\omega_{k_2}}) \right.\nonumber \\
&&\times\left.\int_0^\tau ds\; e^{-i \lambda^{-2}s ({\widetilde\omega}_{k_1}-{\widetilde\omega}_{k_2})} g(s) \right\| \nonumber \\
&<& \epsilon.
\end{eqnarray}
We multiply and divide by $T
|{\widetilde\omega}_{k_1}-{\widetilde\omega}_{k_2}|$ inside the
sum and note that
\begin{equation}
{1\over T |{\widetilde\omega}_{k_1}-{\widetilde\omega}_{k_2}|}
\left|1-
e^{-{T^2({\widetilde\omega}_{k_1}-{\widetilde\omega}_{k_2})^2}}\right|
\leq C <\infty
\end{equation}
is bounded uniformly on $\omega_k$ and $T$ by a constant $C$. This
means that the divergences brought by the rapid $\lambda^{-2}$
oscillations cancel with the slowly varying "spectral average" or
"damping" oscillations, which we treat exactly as in the previous
theorem (see eq.(\ref{eq:analysis}) and the following, also for
the definition of the subsequent $\zeta$) and the norm of the
previous sum is controlled by
\begin{eqnarray}
&&{\lambda^{2-\zeta} C\over T(\lambda)\pi \overline\tau}\!\!\int_0^\infty\!\! dr \!\!\!\!\sum_{k_1,k_2=0}^{N_{\sigma}-1}\!\!\!\! \left\|(E_{\omega_{k_1\!\!+1}}\!\!\!\!\!\!-\!\!E_{\omega_{k_1}}) K_R (E_{\omega_{k_2\!\!+1}}\!\!\!\!\!\!-\!\!E_{\omega_{k_2}}) g(r) \right\| \nonumber \\
&&\leq {\lambda^{2-\zeta} \over T(\lambda)} {C\over \pi}
\|g\|_{-\infty} \| K_R \|
\end{eqnarray}
This shows the uniform boundedness of the last term with respect
to $\|g\|=1$ and $0\leq\tau\leq \overline\tau$, so convergence to
zero comes from the fact that
\begin{equation}
{\lambda^{2-\zeta} \over T(\lambda)}\rightarrow 0 \quad
\lambda\rightarrow 0,
\end{equation}
as we have supposed $T(\lambda)\sim |\lambda|^{-\xi}
\widetilde{T}$ for $\lambda\rightarrow 0$ and $\xi<2$, and we can
take $0<\zeta<2$ such that $2+\zeta-\xi>0$. This finishes the
proof.
\end{proof}

\end{document}